%% file: paper.tex
\documentclass[aps, prb , twocolumn]{revtex4-2}
\usepackage{verbatim}
\usepackage{graphicx}
\usepackage{amsmath}
\usepackage{bbold}
\usepackage{hyperref} 
\usepackage{tikz}
\usepackage[normalem]{ulem}
\hypersetup{ colorlinks=true, linkcolor=blue, citecolor=blue, urlcolor=blue}

\newcommand{\ket}[1]{|#1\rangle}
\newcommand{\bra}[1]{\langle#1|}

\begin{document}
\title{Strong coupling, weak impact: Phonon coupling versus pure dephasing in the photon statistics of cooperative emitters}
\author{J. Wiercinski}
\email[]{jehw2000@hw.ac.uk}
\author{E. Gauger}
\author{M. Cygorek}
\affiliation{SUPA, Institute of Photonics and Quantum Sciences, Heriot-Watt University, Edinburgh EH14 4AS, United Kingdom}
\date{\today}

\begin{abstract}
    Realising scalable quantum networks requires a meticulous level of understanding and mitigating the deleterious effects of decoherence. Many quantum device platforms feature multiple decoherence mechanisms, often with a dominant mechanism seemingly fully masking others.
In this paper, we show how access to weaker dephasing mechanisms can nevertheless be obtained for optically active qubits by performing two-photon coincidence measurements.
To this end we theoretically investigate the impact of different decoherence mechanisms on cooperatively emitting quantum dots. Focusing on the typically dominant deformation-potential coupling to longitudinal acoustic phonons and typically much less severe additional sources of pure dephasing, we employ a numerically exact method to show that these mechanisms lead to very different two-photon coincidence signals. Moreover, surprisingly, the impact of the strongly coupled phonon environment is weak and leads to long-lived coherences. We trace this back to the superohmic nature of the deformation-potential coupling causing inter-emitter coherences to converge to a nonzero value on a short timescale, whereas pure dephasing contributions cause a complete decay of coherence over longer times. 
Our approach provides a practical means of investigating decoherence processes on different timescales in solid state emitters, and thus contributes to understanding and possibly eliminating their detrimental influences.
\end{abstract}

\maketitle
\section{Introduction}
Many quantum technologies crucially rely on scalable quantum networks~\cite{kimble_quantum_2008, ritter_elementary_2012} incorporating several non-classically correlated emitters. These could enable broad applications in quantum computing, quantum communication, quantum metrology, and beyond. Solid-state platforms like nitrogen-vacancy centers \cite{faraon_resonant_2011}, defects in hexagonal Boron Nitride \cite{jungwirth_optical_2017, wigger_phonon-assisted_2019} and self-assembled quantum dots (QDs) \cite{lodahl_interfacing_2015} promise high integrability and stability compared to atomic systems. 

However, due to their inherent interaction with their surrounding environment, all condensed-matter quantum systems are unavoidably subject to decoherence, which is often adequately described by phenomenological pure dephasing (PPD) with a rate determined by experiment. The microscopic origin of such PPD includes charge fluctuations \cite{itakura_dephasing_2003, kuhlmann_charge_2013}, virtual transitions to higher confined states, or higher-order phonon processes~\cite{muljarov_dephasing_2004}.
Yet, in semiconductor nanostructures, the dominant environment effects often stem from the strong coupling to longitudinal acoustic (LA) phonons~\cite{ramsay_damping_2010,ramsay_phonon-induced_2010,quilter_phonon-assisted_2015, koong_coherent_2021}. This coupling can be derived microscopically~\cite{mahan_many-particle_2000,krummheuer_pure_2005}, yielding a strongly frequency-dependent coupling described by a spectral density $J(\omega)$ which approaches zero for $\omega\to0$ as $J(\omega)\propto\omega^3$. We will refer to this coupling as \textit{Superohmic Phonon Coupling} (SPC) for the remainder of this paper~\footnote{In general, a spectral density $J(\omega)$ is called \textit{superohmic}, if $J(\omega)\propto\omega^\nu$, $\nu>1$. This contrasts \textit{subohmic} coupling in the case of $\nu<1$, and \textit{ohmic} coupling in the special case of $\nu=1$.}.
The dynamics resulting from SPC can range from strong non-Markovian behaviour \cite{reiter_distinctive_2019} to situations where polaron \cite{brash_light_2019, koong_fundamental_2019} or Markovian weak coupling Lindblad master equations \cite{ramsay_damping_2010, ramsay_phonon-induced_2010} provide an adequate description. 
On short timescales of the order of picoseconds, SPC dominates the decoherence processes, whereas PPD predominantly affects dephasing on the nanoseconds timescale.
Nevertheless, the impact of both of these sources of decoherence needs to be considered if one desires to push the limits of current light-matter interfaces. 

In this article we show theoretically that it is possible to access decoherence on both timescales by performing two-photon coincidence measurements on two \textit{cooperative emitters}. 
Cooperative emission of an ensemble of resonant emitters differs from the emission of a set of independent emitters by showing, e.g., non-exponential dynamics of the intensity, superextensive scaling of emission rates, or changes in the photon statistics. This is caused by the involvement of inter-emitter coherences in the emission process and typically requires the emitters to be indistiguishable. For solid state systems, spectral indistinguishability has become experimentally achievable due to recent advances in technology allowing for \textit{in situ} control using thermal tuning \cite{kim_super-radiant_2018}, strain \cite{grim_scalable_2019}, or the DC-Stark effect \cite{koong_coherence_2022}. Spatial indistinguishability, however, still proves hard to realize and typically requires placing emitters into specially designed waveguide structures~\cite{kim_super-radiant_2018, grim_scalable_2019}.

It has been shown previously that the measurement of two-photon coincidences from two emitters probes an entangled two-emitter state whose time evolution reflects the dynamics of the inter-emitter coherences that are dominated by the dephasing in the system \cite{cygorek_signatures_2022}. Cooperative emission can be achieved by, e.g., bringing the emitters very close together or erasing information about the source of an emitted photon in the measurement process. These related but distinct cases are known as~ \textit{superradiance} \cite{dicke_coherence_1954} and \textit{measurement-induced cooperativity} \cite{wiegner_quantum-interference-initiated_2011,richter_collective_2022, cygorek_signatures_2022}, respectively, and we will investigate both of them in the present paper.

In the following, we showcase the utility of measuring two-photon coincidences from two cooperative emitters for investigating different dephasing mechanisms for the example of self-assembled GaAs quantum dots (QDs).
Contrasting the influence of SPC non-perturbatively against realistic levels of pure dephasing we find, surprisingly, that on large timescales the strong coupling to longitudinal acoustic phonons is not the main contribution to the outcome of two-photon coincidence measurements which is instead dominated by often neglected dephasing mechanisms. We trace this back to the superohmic coupling spectral density of the deformation-potential coupling that in absence of coherent driving leads only to a partial decay of inter-emitter coherence during a short time interval of few picoseconds. 

Our paper is organized as follows: First, in Section \ref{sec:model}, we introduce our model for the two quantum dots and the measurement as well as our numerical method in subsection \ref{sec:methods}. The following Section \ref{sec:cooperative} then focuses on two-photon coincidence measurements on spatially separated QDs, while treating the superradiant case in Section \ref{sec:superradiance}. Finally, in Section \ref{sec:conclusions}, we summarize our results.

\section{Model}\label{sec:model}
\begin{figure}
    \includegraphics[width=8.6cm]{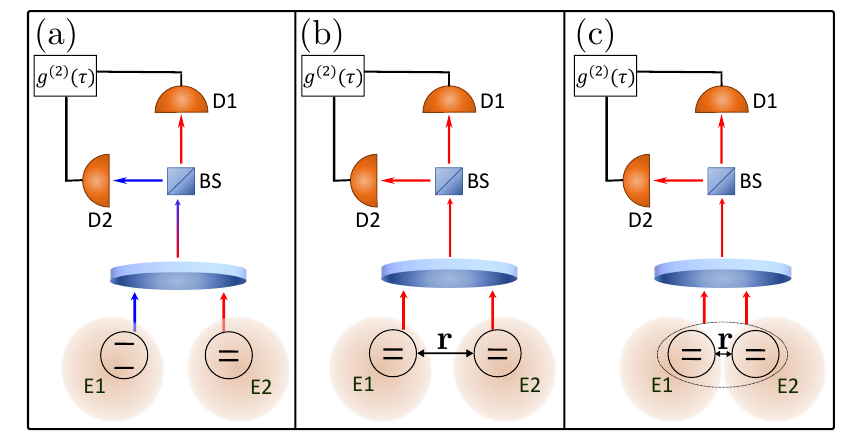}
    \caption{ 
    Two-photon coincidence measurement setups for different configurations of two QDs coupled to two environments E1 and E2. {(a)} Energetically detuned emitters produce photons of different wavelengths (differently coloured arrows). The detection of a photon at detector D1 or D2 determines which QD has decayed (one possible pathway is shown). {(b)} Spectrally indistiguishable emitters at distance $|\textbf{r}|\gg\lambda$. Collecting the emitted light erases information about the source of a measured photon. {(c)} Superradiance: Two resonant emitters at a distance $|\textbf{r}|<<\lambda$ build a combined system (symbolized by the ellipse encircling both emitters) and photons cannot be traced back to a single QD. }\label{fig:g2_measurement}
\end{figure}

We consider two QDs, which we model as two-level systems with ground and excited states $|g_i\rangle$ and $|e_i\rangle$, $i=1,2$, respectively. We denote the corresponding raising and lowering operators by $\sigma_i^+ = |e_i\rangle\langle g_i|$ and $\sigma_i^-= | g_i \rangle\langle e_i\rangle$. A way to observe the photon emission properties of such a system is to perform \textit{two-photon coincidence measurements}, i.e., sending the light emitted by the two QDs into a HBT setup [depicted in Figs.~\ref{fig:g2_measurement}(a)-(c)]. This measurement probes the probability of detecting a photon some time $\tau$ after the detection of a first photon.
\subsection{Two-photon coincidences}
Figure \ref{fig:g2_measurement} depicts three different scenarios:
Spectrally distinguishable QDs, which emit photons independently of each other (cf.~Fig.~\ref{fig:g2_measurement}(a), QDs that are tuned into resonance while spatially separated by distances larger than the wavelength of the emitted light $|\mathbf{r}|\gg \lambda$ [cf.~Fig.~\ref{fig:g2_measurement}(b)], and resonant QDs that are additionally in close proximity to each other with sub-wavelength distance $|\mathbf{r}|\ll\lambda$ [cf.~Fig.~\ref{fig:g2_measurement}(c)] \footnote{In experiments, this can be circumvented by embedding the QDs into a suitable 1d waveguide \cite{kim_super-radiant_2018, grim_scalable_2019}}. The optical beam path is set up such that photons from both QDs are registered at the detectors equally. While photons from distinguishable emitters encode which-path information in the photon frequency, the origin of photons emitted from QDs tuned into resonance cannot be distinguished by the detectors in this setup. As a result, the photon detection is described by a projective measurement with intensity observable $\sigma^+_I \sigma^-_I$, where
$\sigma^\pm_I = \big( e^{\pm i \varphi_1 } \sigma^\pm_1 + e^{\pm i \varphi_2} \sigma^\pm_2 \big)/\sqrt{2}$
with phases $\varphi_i$ generally depending on the optical path length between the $i$-th QD and the detectors.
Without loss of generality, we set $\varphi_1=\varphi_2=0$, as the phases can be absorbed into the definition of the excited states of the respective QDs. Then, we can identify $\sigma^\pm_I=\sigma^\pm_S$, where $\sigma^\pm_S = \big(\sigma^\pm_1 + \sigma^\pm_2 \big)/\sqrt{2}$ is the climbing operator involving the symmetric Dicke state $|\Psi_S\rangle=\big(|e_1,g_2\rangle + |g_1,e_2\rangle\big)/\sqrt{2}$. Therefore, the measured intensity reads \cite{cygorek_signatures_2022}
\begin{align}\label{eq:Intensity}
 I_0= \langle \sigma^+_S\sigma^-_S\rangle = n_{ee} + \frac 12 \big( n_{e,g} +n_{g,e} + c +c^*\big)
\end{align} 
and explicitly depends on the occupations $n_{ee} = \bra{e_1, e_2}\rho\ket{e_1, e_2}$, $n_{e,g} =  \bra{e_1, g_2}\rho\ket{e_1, g_2}$ and $n_{g_1,e_2} =  \bra{g_1, e_2}\rho\ket{g_1, e_2}$ as well as on the inter-emitter coherences $c = \bra{e_1, g_2}\rho\ket{g_1, e_2}$.

Normalised photon coincidences are given by \cite{cygorek_signatures_2022}
\begin{align}\label{eq:two-time_corr}
    g^{(2)}(\tau) &= \lim_{t\to\infty}\frac{\langle\sigma_S^+(t)\sigma_S^+(t+\tau)\sigma_S^-(t+\tau)\sigma_S^-(t)\rangle}{\langle\sigma_S^+(t)\sigma_S^-(t)\rangle\langle\sigma_S^+(t+\tau)\sigma_S^-(t+\tau)\rangle},
\end{align}
where the numerator can be expressed as
\begin{align}
\langle\sigma_S^+(t)\sigma_S^+(t+\tau)\sigma_S^-(t+\tau)\sigma_S^-(t)\rangle
=\langle \sigma_S^+(\tau)\sigma_S^-(\tau)\rangle_{\rho'}
\nonumber\\=
\textrm{Tr}\big[\sigma_S^+(\tau)\sigma_S^-(\tau)\rho'\big],
\end{align} 
where $\rho' = \sigma_S^-(t) \rho \sigma_S^+(t)$ describes the state directly after the projective measurement at the first photon detection.
Defining correspondingly the measurement-induced occupations and coherences as $n'_{ee} = \bra{e_1, e_2}\rho'\ket{e_1, e_2}$, $n'_{e,g} =  \bra{e_1, g_2}\rho'\ket{e_1, g_2}$, $n'_{g_1,e_2} =  \bra{g_1, e_2}\rho'\ket{g_1, e_2}$ and $c' = \bra{e_1, g_2}\rho'\ket{g_1, e_2}$, respectively, the two-photon coincidences take the form 
\begin{align}
\label{eq:g2_explicit}
    g^{(2)}(\tau) =\frac{n'_{ee}(\tau) + \frac{1}{2}\big(n'_{e,g}(\tau) +n'_{g,e}(\tau)\big)}{I_0^2}+
    \frac{\textrm{Re}\big\{c'(\tau)\}}{I_0^2}.
\end{align}
The second part gives a contribution that directly measures the time evolution of measurement-induced coherences. Coherences in the stationary state are reflected in the denominator of Eq.~\eqref{eq:g2_explicit}.

Summarizing, the indinstinguishability of the emitters combined with the equal measurement of the two QDs has two consequences: On the one hand, the projective measurement of the first photon at time $t$ results in the preparation of a correlated state, i.e.~a state with inter-emitter coherences. On the other hand, the detected signal directly probes the time evolution of the measurement-induced coherences. 
It is this dependence of photon coincidences on the measurement-induced coherences that
makes \emph{measurement-induced cooperative emission} an ideal testbed for decoherence in solid-state quantum emitters.
\subsection{Equations of motion}
We consider a system comprised of three parts. First, there is the four-dimensional Hilbert space of the two quantum dots $\mathcal{H}_\text{sys}$. Second, the two quantum dots couple to a continuum of photon modes $\mathcal{H}_\text{phot}$ and, third, to a continuum of phonon modes making up the Hilbert space $\mathcal{H}_\text{phon}$. Correspondingly, we write the total Hamiltonian as
\begin{align}\label{eq:complete_hamiltonian}
    H &= H_{sys} + H_\text{phot} + H_\text{phon} + H_\text{sys-phot} + H_\text{sys-phon}~,
\end{align}
where $H_\text{phot}$ and $ H_\text{phon}$ describe the uncoupled photon and phonon baths, respectively, while $H_\text{sys-phot}$ and $H_\text{sys-phon}$ are the respective Hamilton operators for the coupling of the baths to the two-QD system. We consider the case of degenerate, uncoupled QDs and work in a frame rotating with the transition energy of the QDs, which results in $H_{sys} = 0$. 

Additionally, we assume incoherent pumping of the excited states of the two QDs with identical rates $\gamma_p$ and phenomenological PPD with identical rates $\gamma_d$ for both QDs. In experiments, incoherent pumping can be achieved by coherently driving higher-lying QD-states and relying on incoherent processes to transfer the excitation down to the excited state \cite{koong_coherence_2022}.
Therefore, the evolution of the density matrix describing the combined Hilbert space $\mathcal{H}_\text{sys}\otimes\mathcal{H}_\text{phon}$ reads
\begin{align}
    \dot{\rho} =& \frac{1}{i\hbar}[H_\text{phon}+H_\text{sys-phon}, \rho]\nonumber \\
    &+ \gamma_p(\mathcal{L}_{\sigma_1^+}[\rho]+\mathcal{L}_{\sigma_2^+}[\rho]) + \gamma_d(\mathcal{L}_{\sigma_1^z}[\rho]+\mathcal{L}_{\sigma_2^z}[\rho])\nonumber\\
    &+\mathcal{D}[\rho]\,,\label{eq:time_ev_general}
\end{align}
with the Lindbladian superoperator
\begin{align}
    \mathcal{L}_{O}[\rho] &= O\rho O^\dagger - \frac{1}{2}\left(O^\dagger O\rho + \rho O^\dagger O \right)\label{eq:lindblad}
\end{align} 
and the Markovian radiative decay superoperator $\mathcal{D}$, which we derive in in Sec.~\ref{sec:decay}.
We will further introduce the phonon environment and its coupling to the system in Sec.~\ref{sec:dephasing}. To calculate the reduced density operator $\bar{\rho} = \text{Tr}_{\mathcal{H}_\text{phon}} [\rho]$ as well a two-time correlation functions of reduced system operators we use a numerically exact process tensor method which we present in Sec.~\ref{sec:methods}.
\subsection{Radiative decay}\label{sec:decay}
Both QDs couple to a shared electromagnetic environment. In principle, this environment is comprised of infinitely many modes with wave vector \textbf{k} and polarization $\sigma$. For the purpose of this study,  we suppress the latter without loss of generality. Then, the Hamiltonian for the non-interacting light modes reads
\begin{equation}
    H_\text{phot} = \hbar\sum_\textbf{k}\omega_\textbf{k}a^\dagger_\textbf{k}a_\textbf{k}~,
\end{equation}
where $a_\textbf{k}^\dagger$ ($a_\textbf{k}$) is the creation (annihilation) operator for the photon mode with wave vector $\textbf{k}$. Further,
\begin{align}
    H_\text{sys-phot} &= \hbar\sum_\textbf{k}\left(h_\textbf{k}a_\textbf{k}^\dagger + h_\textbf{k}^\dagger a_\textbf{k}\right)\,,\\
    h_\textbf{k} &= g_\textbf{k}e^{i\textbf{k}\cdot\textbf{r}/2}\sigma_1^- + g_\textbf{k}e^{-i\textbf{k}\cdot\textbf{r}/2}\sigma_2^-\,,
\end{align}
is the light-matter interaction Hamilton operator. In typical scenarios, in which the structure of the photonic modes is not artificially modified, the influence of the photon environment leads to radiative decay that can be adequately captured via Lindblad terms [cf.~Eq.~\eqref{eq:lindblad}]\cite{lindblad_generators_1976}. 
However, their specific form depends on the distance vector $\textbf{r}$ of the QDs.  We here consider two limiting cases: Either the QDs are far apart, i.e.~$|\textbf{r}| \gg \lambda$, where $\lambda$ is the wavelength of the emitted light, or they are very close together, i.e.~$|\textbf{r}| \approx 0$. 
In the case of a large spatial separation the QDs effectively decay independently with rates $\gamma = \frac{2\pi}{\hbar}\sum_\textbf{k}|g_\textbf{k}|^2\delta(\hbar\omega_X-\hbar\omega_\textbf{k})$, where $\omega_X$ is the transition frequency of the QDs. For the remainder of this paper, we assume a realistic single-emitter decay rate of $\gamma^{-1} = 1.75$~ns \cite{koong_coherence_2022}.

Thus, the radiative decay can be described via
\begin{align}\label{eq:master_equation_cooperative}
    \mathcal{D}[\rho] =&  \gamma\left(\mathcal{L}_{\sigma_1^-}[\rho] + \mathcal{L}_{\sigma_2^-}[\rho]\right).
\end{align}
In the case of a vanishing distance vector the dipole operator simplifies to $h_\textbf{k} = \sqrt{2}g_\textbf{g}\sigma_S^-$. This means that the to QDs decay with a collective dipole moment leading to an enhanced decay rate of $2\gamma$:
\begin{align}\label{eq:master_equation_superradiance}
    \mathcal{D}[\rho] = 2\gamma\mathcal{L}_{\sigma_S^-}[\rho]~.
\end{align}
Thus, the emission of a photon from the doubly excited state $\ket{e_1, e_2}$ leads to a transition down the Dicke ladder to the state $\ket{\Psi_S} = \sigma_S^+\ket{g_1, g_2}$. This state is optically bright, while its orthogonal state $\ket{\Psi_A} = \sigma_A^+ \ket{g_1. g_2}$ ($\sigma_A^\pm = \frac{1}{\sqrt{2}}\left(\sigma_1^\pm -\sigma_2^\pm\right)$) is optically dark.
In this decay process the QDs can become entangled via the light-matter interaction.

\subsection{Phonon coupling}\label{sec:dephasing}
In solid-state systems, dephasing, the decay of coherences in quantum states, is a prominent effect. Even though its origins can be diverse, the strong coupling to LA phonons via the deformation-potential coupling is typically assumed to be dominant, especially for InGaAs/GaAs QDs on a short timescale. The corresponding coupling Hamiltonian can be derived microscopically \cite{krummheuer_theory_2002,mahan_many-particle_2000} for one QD. In the case of two QDs the influence of the phonon environment can be captured by considering two separate environments, one coupled to each dot. This is the case because the energy density of a phonon wave packet emitted by one QD decays quadratically with the distance from its origin \cite{wigger_energy_2014} and, in typical experiments, QDs are separated sufficiently far apart for environment-mediated coupling via energy transfer to be negligible.
Moreover, we assume that both QDs are similar in shape and size and surrounding material, allowing us to approximate the two environments to be identical. Consequently, the Hamiltonian for the phonon environment and its interaction with the two QDs reads
\begin{align}
    H_\text{phon} =& \hbar\sum_{i = 1, 2}\left(\sum_\textbf{q}\omega_\textbf{q}b_{i,\textbf{q}}^\dagger b_{i,\textbf{q}}\right)\label{eq:Phonon_Hamilton} ~, \\
    H_\text{sys-phon} =& \hbar\sum_{i = 1, 2}\left(\sum_\textbf{q}g_\textbf{q}\sigma_i^+\sigma_i^-\left(b_{i,\textbf{q}}^\dagger + b_{i,\textbf{q}}\right)\right)
\end{align}
with $g_\textbf{q}$ being the coupling strength of one emitter to the environment mode with wave vector \textbf{q}, $\hbar\omega_\textbf{q}$ the energy of the respective mode and $b^\dagger_{i, \textbf{q}}$ ($b^\dagger_{i, \textbf{q}}$) the creation (annihilation) operator of the $\textbf{q}$-mode of the $i$-th emitter environment.
The influence of the environments on the reduced system can then be fully captured by the spectral density (SD):
\begin{equation}
    J(\omega) = \sum_\textbf{q}|g_\textbf{q}|^2\delta(\omega-\omega_\textbf{q})~.
\end{equation}
The coupling constants $g_\textbf{q}$ can be calculated from the electron and hole wave functions as well as the phonon dispersion relation  $\omega_\textbf{q} = c_s|\textbf{q}|$ of the bulk material, with speed of sound $c_s$ \cite{krummheuer_theory_2002}. Assuming a parabolic confinement potential for electron and hole one arrives at the superohmic spectral density \cite{gauger_high-fidelity_2008} 
\begin{equation}\label{eq:SD_Phonons_Deformation_Potential}
    J_\text{def}(\omega) = \frac{\omega^3}{2\rho \hbar c_s^5}\left(D_ee^{-\frac{\omega^2}{\omega_e^2 }}-D_he^{-\frac{\omega^2}{\omega_h^2 }}\right)^2,
\end{equation}
where $\rho$ is the mass density, $D_e$ ($D_h$) is the electron (hole) deformation potential and $\omega_e$ ($\omega_h$) is the cutoff frequency for electrons (holes) that can be calculated using their effective masses and the confinement strength. Throughout this paper we assume the quantum dots to have a diameter of $4$~nm leading to a cutoff frequencies of $\hbar\omega_e = 2.9$~meV ($\hbar\omega_h = 4.4$~meV). Furthermore, taking InGaAs parameters from Ref.~\cite{krummheuer_pure_2005}, we use $c_s = 5110\text{m}/\text{s}$ for the speed of sound, $\rho = 5370\,\text{kg}/\text{m}^3$ for the mass density and $D_e=7.0\,\text{eV}$ ($D_h=-3.5\text{eV}$) for the electron (hole) deformation potential. The phonon bath temperature is taken to be 4~K.
\subsection{Methods}\label{sec:methods}
\begin{figure}
    \includegraphics[width=8.6cm]{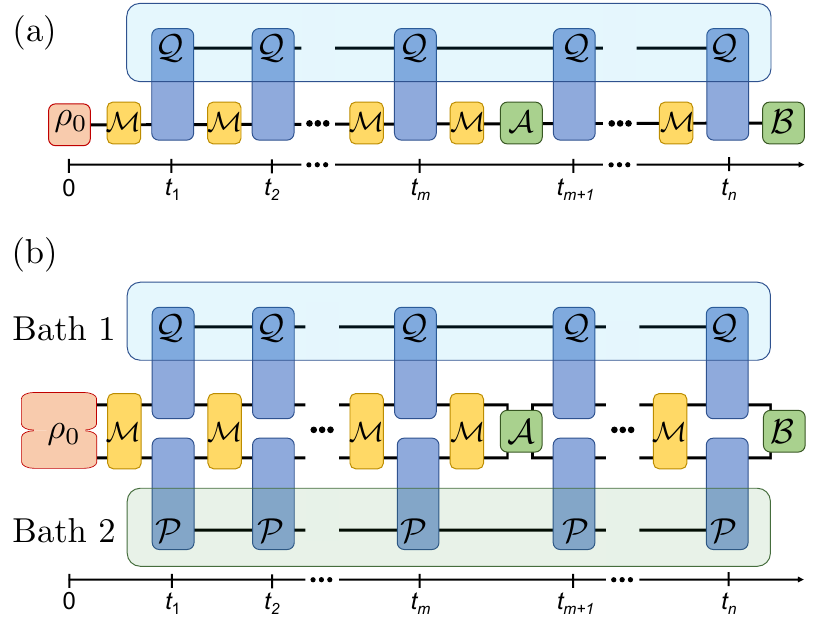}
    \caption{Numerical method to calculate two-time correlation functions: (a) for a single environment: Unitary and Markovian contributions to the time evolution of the reduced system are captured by the tensors $\mathcal{M}$, while the tensors $\mathcal{Q}$ capture the influence of the phonon environment. Information about the environment states is propagated through time by matrix products over the inner indices of the PT. A correlation function $\langle\mathcal{B}(t_n)\mathcal{A}(t_m)\rangle$ can be calculated by inserting the tensor representation of the operators at the corresponding time step. (b) Two-time correlation function for a combined two-emitter system with separate environments. The PTs for the two environments act only on a specific subset of the indices of the whole two-emitter system, while the tensors $\mathcal{M}$ can introduce an effective inter-emitter coupling.}\label{fig:methods}
\end{figure}

The strong electron-phonon interaction in QDs is known to lead to significant non-Markovian memory effects \cite{reiter_distinctive_2019}. These can, however, be described on a numerically exact level using path integral techniques~\cite{barth_path-integral_2016, strathearn_efficient_2017}:
According to Feynman and Vernon \cite{feynman_theory_1963}, the reduced density matrix $\bar{\rho}_{\mu_n \nu_n}(t_n)$ at time step $t_n$ can be calculated as
\begin{align}
\bar{\rho}_{\mu_n \nu_n} =&\sum_{\substack{\mu_{l-1},\dots,\mu_0\\\nu_{l-1},\dots,\nu_0}} \mathcal{I}^{(\mu_n\nu_n)\dots(\mu_1\nu_1)}\bigg(\prod_{l=1}^{n}\mathcal{M}_{\nu_l \nu_{l-1}}^{\mu_l \mu_{l-1}}\bigg)\bar{\rho}_{\mu_0 \nu_0},
 \end{align}
where $\mathcal{M}=e^{\mathcal{L} \Delta t}$ describes the free evolution of the QDs and also includes the Markovian contributions of radiative decay, incoherent pumping and PPD while the influence functional $\mathcal{I}^{(\mu_n\nu_n)\dots(\mu_1\nu_1)}$ fully captures the microscopically modelled phonon effects. 
A process tensor (PT)\cite{pollock_non-markovian_2018} corresponds to an influence functional brought to matrix product operator form \cite{jorgensen_exploiting_2019, cygorek_simulation_2022}
\begin{align}
\mathcal{I}^{(\mu_n\nu_n)\dots(\mu_1\nu_1)}=& \sum_{d_n,\dots,d_0} \bigg(\prod_{l=1}^n\mathcal{Q}^{\mu_l\nu_l}_{d_l d_{l-1}}\bigg).
\end{align}
Its constituents $\mathcal{Q}^{\mu_l\nu_l}_{d_l d_{l-1}}$ are viewed as matrices with respect to the inner indices $d_l$. The role of these indices is to mediate the non-local (non-Markovian) information flow encoded in the influence functional from one time step to the next. This enables a direct time-local propagation of a system coupled to its environment from one time step to the next via
\begin{align}
R_{\mu_l\nu_l}^{d_l}=\mathcal{Q}^{\mu_l\nu_l}_{d_l d_{l-1}}\mathcal{M}_{\nu_l \nu_{l-1}}^{\mu_l \mu_{l-1}}R_{\mu_{l-1}\nu_{l-1}}^{d_{l-1}},
\end{align}
where $R_{\mu_l\nu_l}^{d_l}$ is an extended density matrix with initial state $R_{\mu_0\nu_0}^{d_0}=\rho_{\mu_0\nu_0}\delta_{d_0 1}$. Figure \ref{fig:methods}(a) visually represents the time propagation of the reduced density matrix: The coloured boxes represent the tensors capturing the free system evolution and Lindbladian dissipators (yellow) and the environment influence (blue) at each time step while lines connecting them represent tensor products propagating the information flow. The box encircling the $\mathcal{Q}$-boxes represents the process tensor $\mathcal{I}$.

At intermediate time steps the reduced system density matrix can be obtained via $\bar{\rho}_{\mu_l \nu_l}=\sum_{d_l} q_{d_l}R_{\mu_l\nu_l}^{d_l}$. The closures $q_{d_l}$ can be calculated from the PT using the procedure described in Ref.~\cite{cygorek_simulation_2022}. To calculate two-time correlation functions of two operators $A$ and $B$, one can insert their Liouville space representatives $\mathcal{A}$ and $\mathcal{B}$ into the time evolution at the desired time steps (green boxes in Fig.~\ref{fig:methods}(a)) along the lines of Ref.~\cite{cosacchi_path-integral_2018}. As the process tensor keeps information about the state of the environment intact when the operator $A$ is applied to the system, this procedure remains numerically exact and goes beyond the quantum regression theorem, which can break down in solid-state systems \cite{cosacchi_accuracy_2021}. 

For our purposes, it is straightforward to show that a system in contact with two environments corresponding to PTs with matrices $\mathcal{P}^{\mu_l\nu_l}_{g_l g_{l-1}}$ and $\mathcal{Q}^{\mu_l\nu_l}_{d_l d_{l-1}}$, respectively, can be simulated analogously via
\begin{align}
R_{\mu_l\nu_l}^{g_l d_l }=\mathcal{P}^{\mu_l\nu_l}_{g_l g_{l-1}}\mathcal{Q}^{\mu_l\nu_l}_{d_l d_{l-1}}\mathcal{M}_{\nu_l \nu_{l-1}}^{\mu_l \mu_{l-1}}R_{\mu_{l-1}\nu_{l-1}}^{g_{l-1} d_{l-1}},
\end{align}
where $\bar{\rho}_{\mu_l \nu_l}=\sum_{g_l d_l}q_{g_l} q_{d_l}R_{\mu_l\nu_l}^{g_l d_l }$.
The procedure of including a second environment is still numerically exact because cross-interactions between the environments at time step $l$ are captured in the set of combined indices $(g_l, d_l)$. This very straightforward way of adding environments is a big advantage of process tensor methods over other numerically exact methods like, e.g., hierarchical equations of motion \cite{schinabeck_hierarchical_2018}. In the specific situation presented in this paper, each of the two environments only couples to one of the two emitters. Therefore, it is possible to calculate the process tensor for each of QDs individually \cite{cygorek_code_2021}. Our method for calculating two-time correlation functions involving two process tensors is illustrated graphically in Fig.~\ref{fig:methods}(b). 
\section{Measurement-induced cooperativity of distant QDs}\label{sec:cooperative}
First, we consider the two-photon coincidences from two quantum dots separated by a distance $|\textbf{r}|\gg\lambda$.
According to Eq.~\eqref{eq:g2_explicit}, the photon coincidence signal depends, on the one hand, on the emitted intensity from the stationary state as well as on the occupations and coherences of the state subsequent to the measurement of the first photon. In the case of zero delay time $\tau=0$, the numerator of Eq.~\eqref{eq:two-time_corr} simplifies to the occupation of the double excited state $n_{ee}$ in the stationary state.
In the present case the radiative decay is captured by Eq.~\eqref{eq:master_equation_cooperative} and this stationary state is determined by the balance between pumping and decay for each QD individually. Consequently, for distant emitters no coherences are present in the stationary state, i.e. $c = 0$ [cf.~Eq.~\eqref{eq:Intensity}].
This means that the stationary state is a diagonal product state
\begin{align}
    \rho(t\to\infty) = &n_{gg}\ket{g_1, g_2}\bra{g_1, g_2} + n_{ee}\ket{e_1, e_2}\bra{e_1, e_2}\nonumber\\ &n_{eg}\left(\ket{e_1, g_2}\bra{e_1, g_2}+\ket{g_1, e_2}\bra{g_1, e_2}\right)\,,
\end{align}
with $n_{ee} = n_e^2$ and $n_{eg} = n_en_g$, where $n_e$ ($n_g = 1-n_e$) are the single-emitter occupations of the excited (ground) state.  Using the above considerations and inspecting Eq.~\eqref{eq:g2_explicit}, one finds that $g^{(2)}(0)=1$. Even more surprisingly, in the absence of dephasing,  $g^{(2)}(\tau)=1$ for all delay times $\tau$  \cite{cygorek_signatures_2022}. 

\begin{figure*}[ht]
    \includegraphics[width=17.2cm]{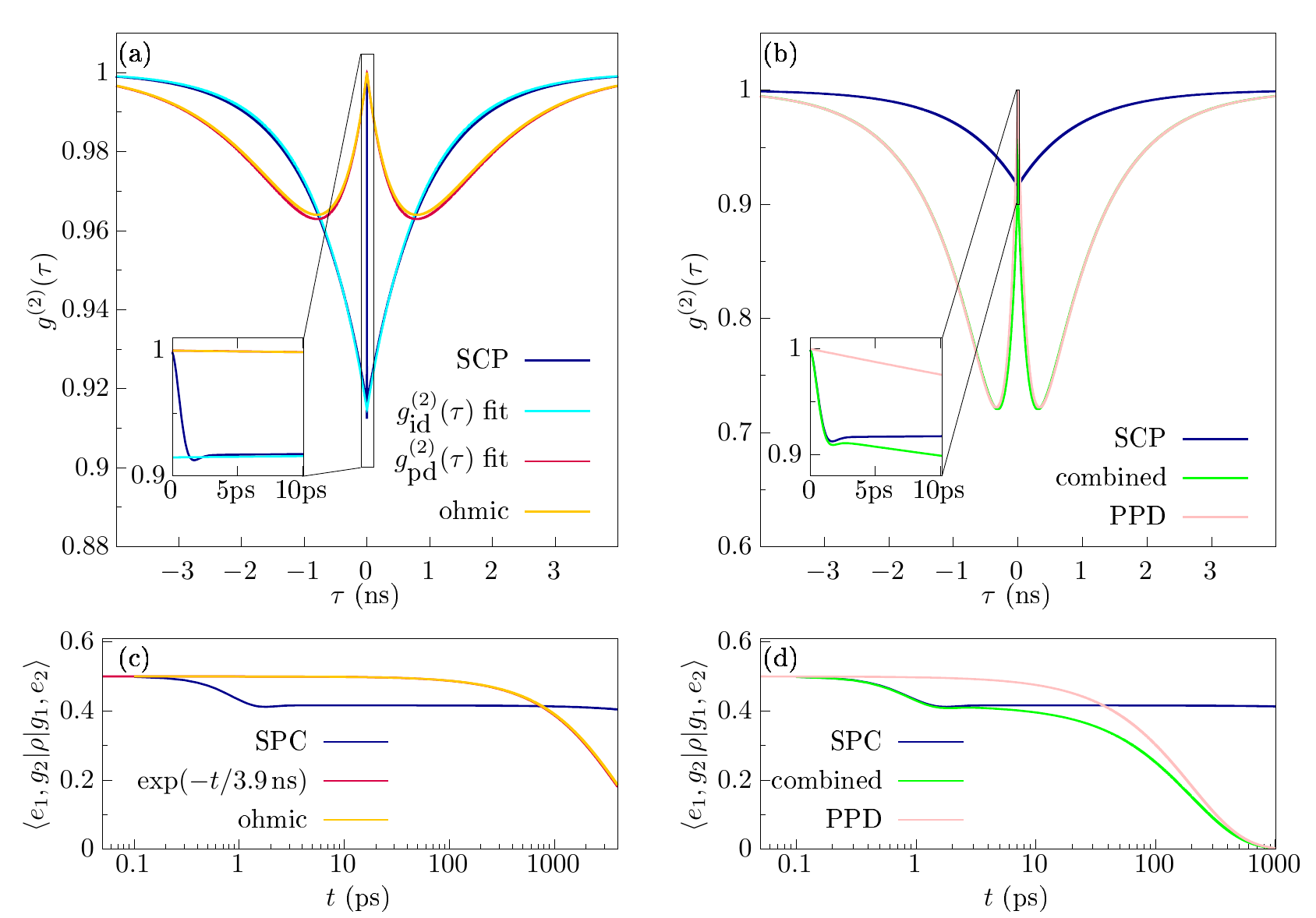}
    \caption{(a) Two-photon coincidences from spatially separated emitters with radiative decay rate $\gamma^{-1} = 1.76$~ns for SPC (blue) and fits of Eq.~\eqref{eq:phonon_g2_approximation} (cyan) and the PPD model Eq.~\eqref{eq:g2_dephasing_rate} (red). The values obtained from fits are $a = 0.0854$ for $g^{(2)}_\text{id}(\tau)$ and $\gamma_d^{-1} = 3.9$~ns for $g^{(2)}_\text{pd}(\tau)$. The inset provides a closer look into the region around $\tau\approx 0$. Additionally, the two-photon coincidences for an ohmic spectral density are shown. b) $g^{(2)}(\tau)$ for SPC (blue), SPC with additional PPD with rate $\gamma_d^{-1} = 221$~ps (green line) and PPD with a rate of $\gamma_d^{-1} = 0.199$~ps (pink) consistent with experiments\cite{koong_coherence_2022}. The region $\tau\approx0$ is shown in the inset. (c) Inter-emitter coherences of a two-emitter system initially prepared in the Dicke state $\ket{\Psi_S}$ for the types of dephasing considered in (a): SPC (blue line), PPD with rate $\gamma_d^{-1} = 3.9$ ns (red), ohmic (yellow) a. Note the logarithmic scale on the time axis. (d) Inter-emitter coherences of a two-emitter system initially prepared in the Dicke state $\ket{\Psi_S}$ for the types of dephasing considered in (c): SPC (blue line), combined SPC and PPD (green line ) and PPD (pink line)}\label{fig:cooperative_fig_1}
\end{figure*}

We first investigate the impact of SPC on the delay time dependence of the photon coincidences, which is depicted in Fig.~\ref{fig:cooperative_fig_1}(a) for $\gamma_p^{-1}=\gamma^{-1}=1.76$~ns. We find that $g^{(2)}(\tau)$ drops by about $10\%$ very quickly and then approaches unity for $\tau\to\infty$. A closer look into the region $\tau\approx0$ [cf.~inset in Fig.~\ref{fig:cooperative_fig_1}(a)] reveals that this drop happens in about one to two picoseconds. To further analyse the behaviour we consider two ad hoc models and fit them to the numerically exactly calculated $g^{(2)}(\tau)$ function. First, we check if the impact of SPC can be emulated by a PPD model with an adapted rate. For PPD with $\gamma_p=\gamma$, $g^{(2)}(\tau)$ reads \cite{cygorek_signatures_2022} 
\begin{equation}
    g_\text{pd}^{(2)}(\tau) = 1-\frac{1}{2}\left(e^{-2(\gamma+\gamma_p)|\tau|} - e^{-(\gamma+\gamma_p+\gamma_d)|\tau|}\right).\label{eq:g2_dephasing_rate}
\end{equation}
Performing a least-square fit and comparing the result to the two-photon coincidences due to SPC we see that PPD captures SPC very badly [cf. Fig.\ref{fig:cooperative_fig_1}(a)]: It neither covers the short timescale depicted in the inset nor the large-scale behaviour. Assuming, however, that the influence of SPC only leads to an initial drop and can be neglected afterwards, we fit 
\begin{equation}\label{eq:phonon_g2_approximation}
    g_\text{id}^{(2)}(\tau) = 1-ae^{-(\gamma+\gamma_p)|\tau|}\,,
\end{equation}
where $a$ is given by the initial decay of $g^{(2)}(\tau)$. Indeed, this closely captures the behaviour of the photon coincidences outwith the region very near $\tau=0$ [cf.~Fig.~\ref{fig:cooperative_fig_1}(a)].

The fact that SPC only influences the photon emission of this system on short timescales is due to the lack of coherent driving. Then, the SPC leads to the formation of a polaron accompanied by a loss of inter-emitter coherences. For low temperatures, however, this loss is finite~\cite{krummheuer_theory_2002}. Therefore, coherences remain in the system for times much longer than typical phonon time scales of a few picoseconds, before they are eventually destroyed by the incoherent pumping and radiative decay. 

The finite long-time coherences are a direct consequence of the superohmic shape of the deformation-potential spectral density \eqref{eq:SD_Phonons_Deformation_Potential} and have also been discussed for ground-to-excited state coherences in single QDs~\cite{krummheuer_theory_2002} which are captured by the \textit{independent boson model}~\cite{breuer_theory_2007}. We show in  App.~\ref{app:mapping}, that the same reasoning as for one QD can be applied to the case of two QDs with the inter-emitter coherences taking the role of the coherences in the independent boson model. This is because for low pumping and deacy rates $\gamma_p$ and $\gamma$, respectively, only the single-excitation subspace contributes noticeably to the emission of the second photon in the two-photon conicidence measurement and this space can be mapped onto an independent boson model.

Figure \ref{fig:cooperative_fig_1}(c) contrasts the different dephasing influences of PPD and SPC: Starting from the initial state $\ket{\Psi_S}$, it shows the inter-emitter coherences in absence of pumping and radiative decay. PPD, on one hand, describes an exponential decay of coherences on a timescale of nanoseconds (red line). In contrast, SPC leads to an initial decrease of inter-emitter coherences in a few picoseconds, but does not reduce them to zero. 

The differences of impact in the SPC and the PPD case do not stem from the approxiations made to arrive at a Lindblad description of the dephasing. On the contrary, the fact that the coherences do not decay for SCP is completely due to the shape of the superohmic spectral density. To show this, we calculate the two-photon coincidences for an ohmic spectral density $J(\omega) = \alpha\omega\exp(-\omega^2/\omega_c^2)$ with $\alpha = 7.5\cdot10^{-5}$ and $\hbar\omega_c = 4$~meV. The cutoff frequency has been chosen to the mean value between the electron and hole cutoff frequencies used for the SPC.
Fig.~\ref{fig:cooperative_fig_1}(a) shows that in this case the two-photon coincidences are captured very well by PPD, as well as the coherences [cf.~Fig.\ref{fig:cooperative_fig_1}(c)]. 

After having established that the impact of SPC on two-photon coincidence measurements on two incoherently driven QDs is limited to the time of polaron formation we now turn to a realistic experimental situation. For real systems the strong coupling to longitudinal acoustic phonons is typically dominant, but other dephasing mechanisms exist as well, which are typically of the PPD-type. To investigate the influence of these additional sources of dephasing, we perform calculations with a realistic PPD rate added to the SCP influence. The results are shown in Fig.~\ref{fig:cooperative_fig_1}(b): The phonon influence is completely masked by the PPD contribution. Considering that in almost all cases SPC is known to be the dominant dephasing mechanism this is a striking result. This seeming contradiction can be resolved when considering the inset of Fig.~\ref{fig:cooperative_fig_1}(b). On short timescales the SPC contribution dominates over the PPD contribution. It is just the absence of coherent driving that restricts its influence to short times while the long timescales are determined by the PPD contributions. Comparing this with recent experimental findings \cite{koong_coherence_2022}, we show in Fig.~\ref{fig:cooperative_fig_1}(b) that the PPD model used for describing experimental data can be reproduced by considering combined SPC and comparatively weak PPD. 

Figure \ref{fig:cooperative_fig_1}(d) Depicts the coherences in the absence of pumping and decay in this situation: In reality we expect a combination of SCP with additional PPD contributions that features a fast initial drop due to SCP and afterwards a slow exponential decay of the coherences due to PPD. 

It can be seen from Figs.~\ref{fig:cooperative_fig_1}(b) and (d) that the photon coincidence measurement separates the timescales of both of these processes and makes both of them -- in principle -- observable independently of each other. In a realistic experiment, however, finite instrument resolution limits the ability to resolve the initial drop due to SPC. We show in App.~\ref{app:convolution} that this leads to the SPC to be observable as a reduced value of $g^{(2)}(0)$.

As the key result of this study, we have found that in the case of measurement-induced cooperative emission the outcome of two-photon coincidence measurements most strongly depends on slow pure dephasing as opposed to the usually dominant SPC. This allows experimental access to typically neglected contributions to decoherence.

\section{Superradiant QDs}\label{sec:superradiance}
In contrast to the previously discussed case of measurement-induced cooperative emission, the superradiant decay process of two very close, identical QDs involves, by its nature, inter-emitter coherences. In the master equation \eqref{eq:master_equation_superradiance} this is reflected in the Lindblad operator $\mathcal{L}_{\sigma_S^-}$ which describes transitions through the maximally entangled symmetric Dicke state $\ket{\Psi_S}$. This leads to correlations in the steady state and impacts the value $g^{(2)}(0)$. We want to discuss the dependence of $g^{(2)}(0)$ on the pumping strength and decay rate first, before turning to the impact of SPC and PPD on the delay-time dependence.

\begin{figure*}
    \includegraphics[width=17.2cm]{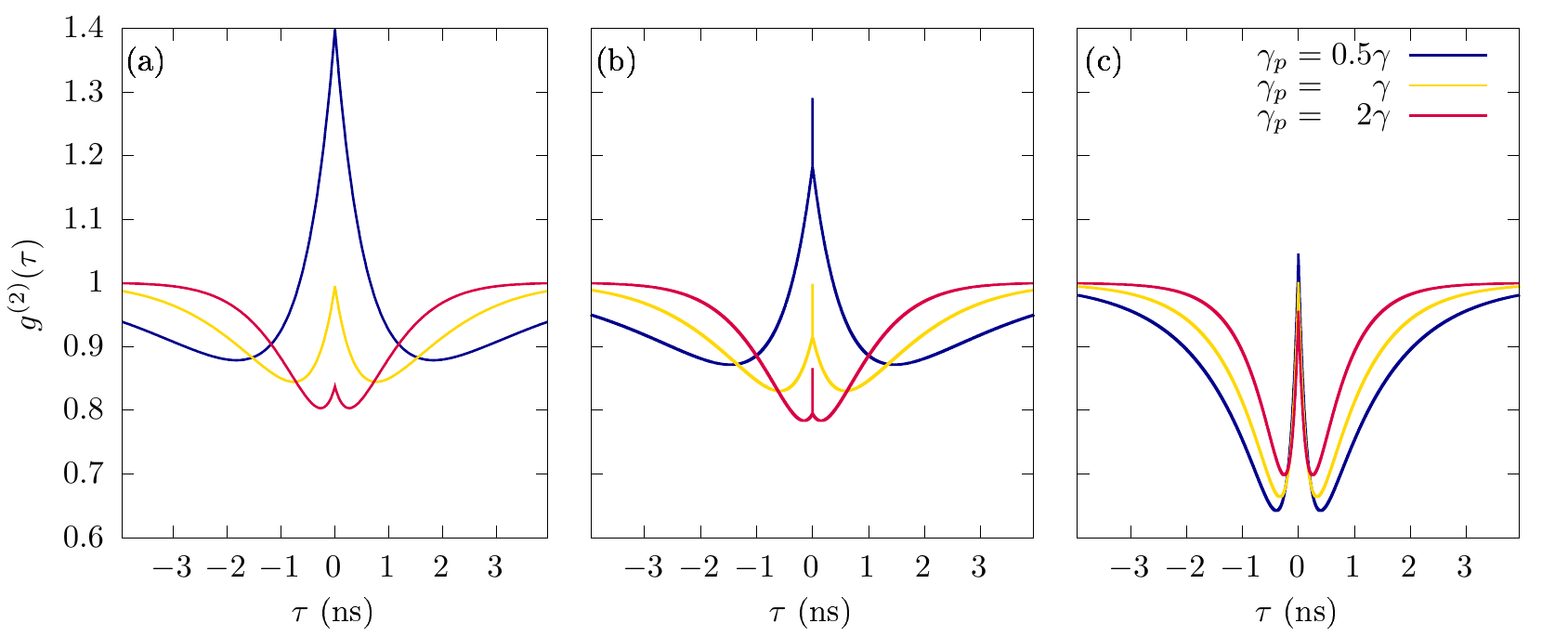}
    \caption{Two-photon coincidences for superradiant emitters for different pumping rates. (a) In the case of no dephasing, (b) for SPC, (c) for a PPD with rate $\gamma_d^{-1} = 199$ ps. The decay rate of the individual QDs is $\gamma^{-1} = 1.76$ ns across all panels.}\label{fig:superradiance_fig_1}
\end{figure*}

Fig.~\ref{fig:superradiance_fig_1}(a) shows, that in absence of dephasing $g^{(2)}(0)$ can take values above one. More precisely, for $\gamma_p/\gamma < 1$ $g^{(2)}(0)>1$, while $\gamma_p/\gamma > 1$ $g^{(2)}(0)<1$, with the special case $\gamma_p=\gamma$ giving $g^{(2)}(0)=1$. 
This can be understood by casting Eq.~\eqref{eq:two-time_corr} into the form $g^{(2)}(0) = \frac{n_{ee}}{(n_{ee} + n_S)^2}$, where $n_S$ is the occupation of the symmetric Dicke state $\ket{\Psi_S}$.
For $\gamma_p/\gamma < 1$ the occupation of $\ket{\Psi_S}$ is lower than the occupation of the antisymmetric dark state $\ket{\Psi_A}$. Leading to negative coherences, this reduces the denominator. Conversely for $\gamma_p/\gamma > 1$, the occupation of $\ket{\Psi_S}$ exceeds the occupation of $\ket{\Psi_A}$ and thus $g^{(2)}(0)<1$.

Before turning to the full dependence on the delay time $\tau$, we investigate how dephasing impacts the zero-delay photon coincidences $g^{(2)}(0)$, which reflects the impact of the dephasing mechanisms on the stationary state of the system. From Fig.~\ref{fig:superradiance_fig_1}(b) and Fig.~\ref{fig:superradiance_fig_1}(c) one finds that SPC as well as PPD do not change which state, $\ket{\Psi_S}$ or $\ket{\Psi_A}$, is occupied predominantly, but rather tends to balance the occupations of the two states and therefore reduces the modulus of the coherence, bringing to $g^{(2)}(0)$ closer to unity. However, one clearly finds differences in strength of impact and a PPD rate of $\gamma_d\approx 10\gamma$ leads to both states being almost evenly occupied, while SPC redistributes the occupations far less efficiently.

Having understood the zero-delay two-photon coincidences, we now turn to their delay-time dependence, which is depicted in Fig.~\ref{fig:superradiance_fig_1}. First of all we find that the impacts of SPC and PPD, as shown in Figs.~\ref{fig:superradiance_fig_1}(b) and \ref{fig:superradiance_fig_1}(c), respectively, are qualitatively different.

Like in the case of measurement-induced cooperative emission, SPC leads to a fast initial drop on the timescale of the polaron formation. Apart from this initial decay the SPC strongly resembles the dephasing-free case. Applying the same reasoning as in the previous section, we conclude that this is because, except for the region around $\tau=0$, SPC does not introduce dephasing due to the lack of coherent pumping. Indeed, taking into account a finite time resolution, we do not expect the initial phonon-induced decay to be observable. As shown in App.~\ref{app:convolution}, the only visible difference to a dephasing-free case is a reduced value of $g^{(2)}(\tau)$.

When looking at the PPD results one notices that compared to the SPC and dephasing-free results the two-photon coincidences possess a pronounced anti-dip whose depth decreases, when increasing the pumping strength. However, this dependence is relatively weak compared to the strong impact $\gamma_p$ has in the dephasing-free case. One can trace back this relatively weak dependence on $\gamma_p$ to $\gamma_d$ being significantly larger than $\gamma_p$ and $\gamma$. The minor changes in depth and width of the anti-dip are due to the fact that with increasing driving strength, the recovery to the stationary state is faster. Compare this again to SPC: Due to limited impact of SPC, the dependence on $\gamma_p$ is far more pronounced.

\begin{figure}
    \centering
    \includegraphics[width=\linewidth]{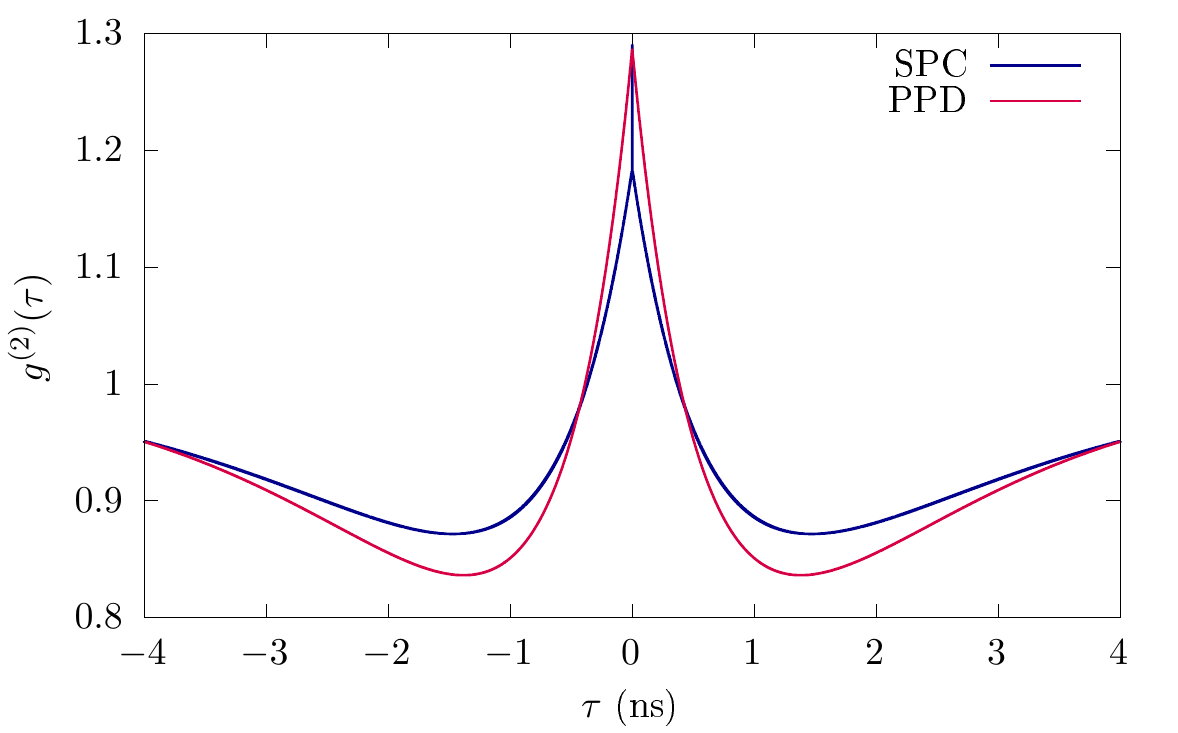}
    \caption{PPD approximation to the SCP in the case of superradiant emitters. The pure dephasing rate has been determined to be $\gamma_d^{-1} = 3.9$ ns in the previous section. The individual emitter decay rate is $\gamma^{-1}=(2\gamma_p^{-1}) = 1.76$ ns.}\label{fig:superradiance_fig_2}
\end{figure}

However, until now, much of the influence of PPD has been traced back to the dephasing rate being large compared to the pumping and decay rates. Therefore, we choose the very small dephasing rate previously extracted via a least squares fit in section \ref{sec:cooperative} and compare the resulting two-photon coincidences with the SPC case in Fig.~\ref{fig:superradiance_fig_2}. Two observations can be made: First of all, the initial values $g^{(2)}(0)$ are approximately the same in both cases. This is in agreement with the PPD approximation to SPC in Sec.~\ref{sec:cooperative}. Additionally, the difference between those cases is relatively minor, especially if compared to the difference of influence of SPC and PPD on cooperative emission. This is the case because the interplay between incoherent pumping and the superradiant decay mechanism strongly contributes to the time evolution of the coherences and therefore, by virtue of Eq.~\eqref{eq:g2_explicit}, the shape of the two-photon coincidences in the superradiant case. In contrast, in the case of measurement-induced cooperative emission of spatially separated QDs, the coherences, and thereby $g^{(2)}(\tau)$, are solely determined by the dephasing mechanism. Thus the superradiant decay masks the differences between the different types of dephasing.
\section{Conclusions}\label{sec:conclusions}
We have used state-of-the-art open quantum system modelling to investigate the influence of different dephasing mechanisms on cooperative emission from two resonant quantum dots, each coupled to a local environment.
Two distinct situations have been considered. First, we investigated two quantum dots that are far apart, where the emission shows clear signs of collective behavior due to the measurement-induced preparation of an Dicke state with inter-emitter correlations. Additionally, we considered superradiant quantum dots, which are very close together and consequently decay collectively, accompanied by an enhancement of the decay rate. 

Focusing on modeling two-photon coincidence measurements we found that the influence of longitudinal acoustic phonons due to deformation-potential coupling is very distinct to that caused by pure dephasing. This is due to the fact that in the absence of coherent driving the inter-emitter coherences decay only partially on a very short timescale. In presence of realistic dephasing rates, on the other hand, pure dephasing leads to a complete decay of the coherences on a timescale comparable with radiative decay. 
This leads to the surprising observation that the shape of two-photon coincidence signals of cooperatively emitting quantum dots is only slightly affected by the strong coupling to longitudinal acoustic phonons, while it is strongly affected by seemingly weak pure dephasing. We showed that pure dephasing as well as ohmic environments are suitable candidates for explaining recent experimental observations \cite{koong_coherence_2022}. Possible sources for such additional relatively mild dephasing include higher-order phonon contributions due to higher-lying quantum dot states \cite{muljarov_dephasing_2004}, charge-carrier fluctuations \cite{itakura_dephasing_2003}, or fluctuations of the applied electromagnetic fields~\cite{volker_dephasing_2000}.

Comparing the case of superradiance and measurement-induced cooperativity we find that, in the latter case, the time dependence of the inter-emitter coherences is relatively easily accessible, while their impact is harder to see in the superradiant case due to an increased number of competing influences which all substantially impact emitter coherences rendering the specific contribution attributable to slow dephasing processes less clear.

In conclusion, we propose that careful investigations of two-photon coincidences in solid-state emitters can contribute to the understanding of not only dominant dephasing effects but also other decoherence influences that are typically masked and hard to access.

\appendix
\section{Mapping of the two-emitter problem to the independent boson model}\label{app:mapping}
In the main text we liken the behavior of the two QDs to the independent boson model. The latter describes a two-level system coupled to a continuum of boson modes. Here we show that the two-emitter-two-environments problem can be mapped to the problem of a single two-level system coupled to one boson bath with the same spectral density, if one restricts oneself to the single-excitation manifold of the two QDs. This, however, is sufficient because, if the pumping is sufficiently weak, two photons can only be emitted if the emission of the first photon is caused by a transition $\ket{e_1, e_2}\to\ket{\Psi_S}$. Subsequent emission of a photon then depends on the occupation of $\ket{\Psi_S}$, which can only change due to dephasing-induced transitions $\ket{\Psi_S}\to\ket{\Psi_A}$. Consequently, dephasing within the single-excitation manifold dominates the time-delay dependence of the photon coincidences. 

Considering the Hamilton operator \eqref{eq:Phonon_Hamilton}, one can introduce symmetrised operators 
\begin{align}
    b_{S,\textbf{q}} = \frac{1}{\sqrt{2}}\left(b_{1,\textbf{q}} + b_{2,\textbf{q}}\right)\,, \\
    b_{A,\textbf{q}} = \frac{1}{\sqrt{2}}\left(b_{1,\textbf{q}} - b_{2,\textbf{q}}\right)\,,
\end{align}
which obey the Bose commutation relations
\begin{align}
    \left[b_{S,\textbf{q}}, b_{S,\textbf{q}^\prime}^\dagger\right] = \delta_{\textbf{q}\textbf{q}^\prime}\,,\\
    \left[b_{A,\textbf{q}}, b_{A, \textbf{q}^\prime}^\dagger\right] = \delta_{\textbf{q}\textbf{q}^\prime}\,.
\end{align}
Thus, one can rewrite the Hamilton operator:
\begin{align}
    &H_\text{phon}+H_\text{sys-phon} \nonumber\\
     &= \hbar\sum_\textbf{q}\omega_\textbf{q}(b_{S,\textbf{q}}^\dagger b_{S,\textbf{q}} + b_{A,\textbf{q}}^\dagger b_{A,\textbf{q}})\nonumber\\ &+ \left(\sigma_1^+\sigma_1^-+\sigma_2^+\sigma_2^-\right)\sum_\textbf{q}\frac{\hbar g_\textbf{q}}{\sqrt{2}}\left(b_{S,\textbf{q}}^\dagger + b_{S,\textbf{q}}\right)  \nonumber\\&+ \left(\sigma_1^+\sigma_1^--\sigma_2^+\sigma_2^-\right)\sum_\textbf{q}\frac{\hbar g_\textbf{q}}{\sqrt{2}}\left(b_{A,\textbf{q}}^\dagger + b_{A,\textbf{q}}\right)
\end{align}
Projecting the Hamiltonian onto the single-excitation subspace in the basis $\{\ket{g_1, e_2}, \ket{e_1, g_2}\}$, $(\sigma_1^+\sigma_1^- + \sigma_2^+\sigma_2^-)$ reduces to the identity, while $(\sigma_1^+\sigma_1^- - \sigma_2^+\sigma_2^-)$ reduces to $\sigma_z$.
Thus, ignoring the symmetric modes that decouple from the system, one arrives at an effective Hamiltonian
\begin{equation}
    H_{SE} = \hbar\sum_\textbf{q}\omega_\textbf{q}b_{A,\textbf{q}}^\dagger b_{A,\textbf{q}} + \frac{\sigma_z}{2}\sum_\textbf{q}\sqrt{2}\hbar g_\textbf{q}\left(b_{A,\textbf{q}}^\dagger + b_{A,\textbf{q}}\right),
\end{equation}
which describes an independent boson model \cite{breuer_theory_2007} with spectral density:
\begin{equation}
    J_{SE} = \sum_\textbf{q}|\sqrt{2}g_\textbf{q}|^2\delta(\omega-\omega_\textbf{q}).
\end{equation}
The factor of two appears because the decoherence effects of both phonon baths add up. 

Summarizing, we indeed find that the single-excitation subspace of the two-emitter problem with two identical, but separate, baths reduces to an independent boson model with double the spectral density of one of the baths. 

\section{Results for a realistic instrument response times}\label{app:convolution}
It has been shown in the main text that the influence of SPC in two-photon coincidence measurements is restricted to very short times compared to the typical timescale of radiative decay. Especially the initial drop of $g^{(2)}(\tau)$ is well beyond typical instrument resolution \cite{natarajan_superconducting_2012}. 

In this appendix we discuss what one can expect the two-photon coincidences to look like in a typical experimental realization of the experiment by taking into account a finite instrument response time. For this we assume the instrument response function to be well-approximated by a Gaussian of FWHM$\approx 240$~ps [cf.~Ref.~\cite{koong_coherence_2022}] and perform a convolution with this instrument response with our results for the two-photon coincidences. 
\begin{figure*}
\centering
    \includegraphics[width=12.9cm]{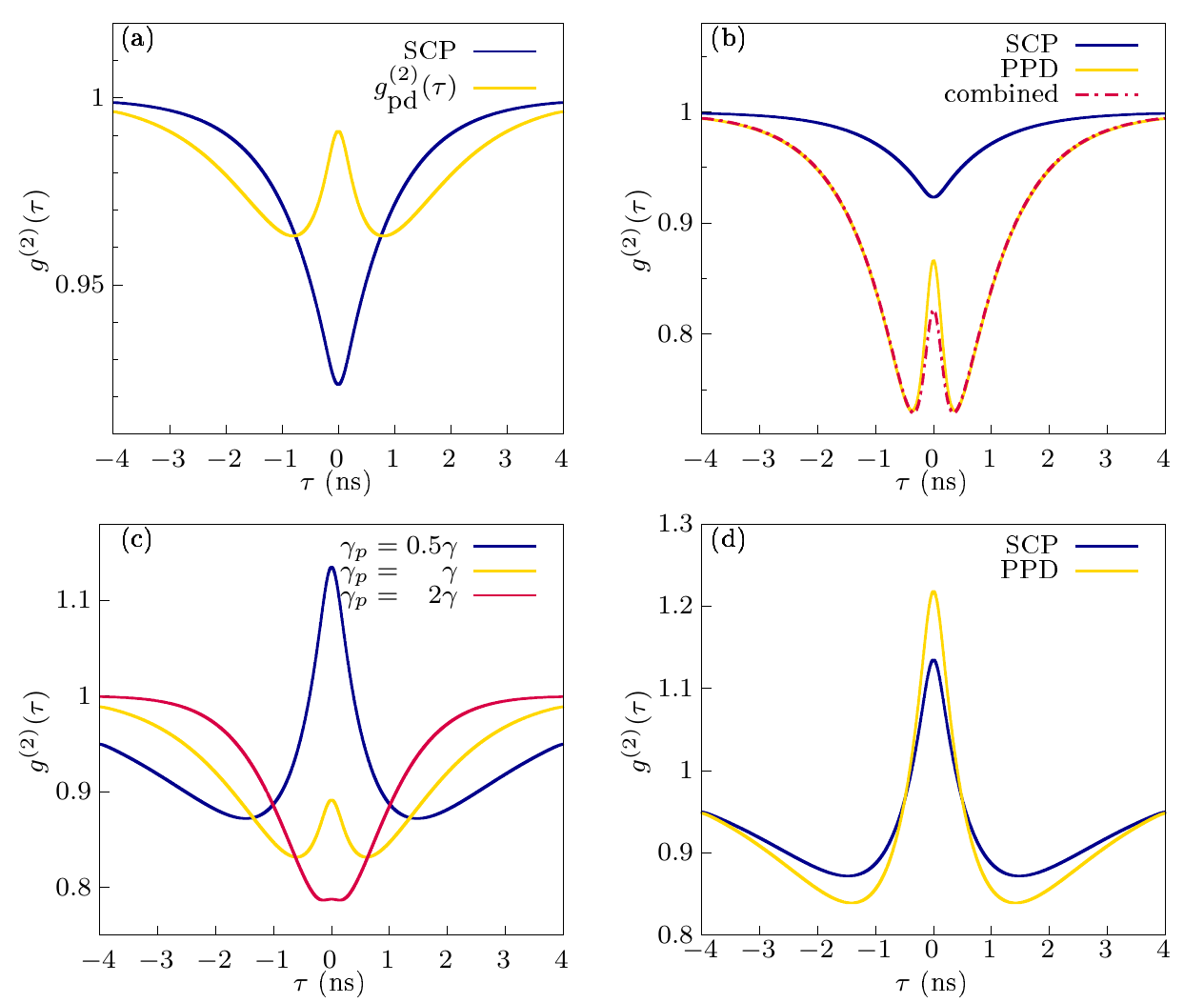}
    \caption{Two-photon coincidence results for a finite instrument resolution of $\approx 240$ps. (a) Two-photon coincidences for SCP and PPD approximation, for the parameters we refer to Fig.~\ref{fig:cooperative_fig_1}(a); (b) Two photon coincidences for SCP, PPD with an expperimentally obtained rate and a combined SCP-plus-PPD model that reproduces the experiment [cf. Fig.~\ref{fig:cooperative_fig_1}(b) for parameters]. (c) Analog to Fig.~\ref{fig:superradiance_fig_1}(a); and (d) analog to Fig.~\ref{fig:superradiance_fig_2}}\label{fig:convolution_fig_1}
\end{figure*}

Figure \ref{fig:convolution_fig_1}(a) shows the two photon coincidences due to SCP and the best PPD approximation, in analogy to Fig.~\ref{fig:cooperative_fig_1}(a), for a finite instrument response time. Looking at the SPC results one finds indeed that the anti-dip completely vanishes and instead the value measured for zero delay time is reduced. On the other hand, the anti-dip vanishing  for SPC means that the absence of an anti-dip cannot unambigously be used to infer absence of cooperativity. The cooperative character of two emitters is rather reflected by $g^{(2)}(0)>0.5$. Compare this to the PPD approximation. In this case, even though the two-photon coincidences do not approach unity for zero delay time, one still finds remains of the anti-dip, though less pronounced. This means that, no matter the finite instrument response, SPC can be very well distinguished from PPD.

In Fig.~\ref{fig:convolution_fig_1}(b) we show SCP, the PPD approximation to experimental data [cf.~Ref.~\cite{koong_coherence_2022}], and the combined model that reproduces the latter one, like in Fig.~\ref{fig:cooperative_fig_1}(b). While the measured signals of PPD and the combined model are similar for $\tau>0.5$~ns, the SCP contribution leads to a decrease of the measured zero-delay two-photon coincidences in the combined case compared to PPD.

Turning now to superradiance, the initial drop of $g^{(2)}(\tau)$ due to SPC cannot be resolved, but the value of $g^{(2)}(0)$ is reduced compared to the non-convoluted results [cf.~Figs.~\ref{fig:convolution_fig_1}(c) and \ref{fig:superradiance_fig_1}(a)]. The anti-dip, however, still survives for small pumping strengths because in the superradiance case it is not caused by the dephasing process alone. At last consider Fig.~\ref{fig:convolution_fig_1} (d). For superradiant emitters, the pure dephasing approximation does reasonably well to describe the phonon influence. Thus, we do not expect that one can clearly distinguish phonon effects from pure dephasing for superradiant emitters, while this is certainly true -- as we have seen in Fig.~ \ref{fig:convolution_fig_1}(a) -- for emitters in the scenario of measurement-induced cooperative emission.
\bibliographystyle{apsrev4-2}
\input{paper.bbl}
\end{document}

%% file: paper.bbl
%